\documentclass[iop]{emulateapj}

\usepackage{graphicx}
\usepackage{color}
\usepackage{amsmath,amssymb}
\usepackage{type1cm}
\usepackage{ulem}
\usepackage{mathrsfs}

\slugcomment{draft version \today, }

\shorttitle{Emission From Dead Radio-Loud Active Galactic Nuclei}                                                      
\shortauthors{Ito et al.}

\begin{document}

\title{The Fate of  Dead Radio-loud Active Galactic Nuclei: A \\ 
New Prediction of Long-lived Shell Emission}

\author{Hirotaka Ito\altaffilmark{1}, Motoki Kino\altaffilmark{2}, Nozomu Kawakatu\altaffilmark{3}, Monica Orienti\altaffilmark{4}} 

\altaffiltext{1}{Astrophysical Big Bang Laboratory, RIKEN, Saitama 351-0198, Japan}
\altaffiltext{2}{Korea Astronomy and Space Science Institute,  776 Daedukdae-ro, Yuseong-gu, Daejeon 305-348, Korea}
\altaffiltext{3}{National Institute of Technology, Kure, 2-2-11 Agaminami,
Kure, Hiroshima, 737-8506, Japan}
\altaffiltext{4}{INAF Istituto di Radioastronomia, via Gobetti 101, I-40129 Bologna, Italy}

\email{hirotaka.ito@riken.jp}

\begin{abstract}

 We examine the fate of a dead radio source
 in which  jet injection from the central engine has stopped
 at an early stage of its evolution ($t = t_{\rm j} \lesssim 10^5~{\rm yr}$). 
 To this aim, we theoretically evaluate the evolution of 
 the emission from both the
 lobe and the shell, which are composed of shocked jet matter and 
 a shocked ambient medium (i.e., shell), respectively.
 Based on a simple dynamical model of expanding lobe and shell,
 we clarify how the broadband spectrum of each component
 evolves before and after the  cessation of the jet activity.
It is shown that the spectrum is strongly dominated by the lobe emission
while the jet is active ($t\leq t_{\rm j}$).
On the other hand, once the jet activity has ceased ($t > t_{\rm j}$),
the lobe emission fades out rapidly,  since fresh electrons are no longer supplied from the jet.
 Meanwhile, shell emission only shows a gradual decrease, 
 since fresh electrons are continuously supplied from the
 bow shock that is propagating into the ambient medium.
 As a result,
 overall
 emission from  the shell
 overwhelms that from the lobe 
 at  wide range of frequencies from radio up to
 gamma-ray soon after the jet activity has ceased.
 Our result predicts a new class of
 dead radio sources that are dominated by  shell emission. 
 We suggest that the  emission from the shell
 can be probed in particular at a radio wavelengths with the Square Kilometer Array (SKA) phase 1.

\end{abstract}


\section{INTRODUCTION}

 Relativistic jets in radio-loud active galactic nuclei (AGNs) 
 dissipate their kinetic energy  via 
 interactions with surrounding 
 interstellar medium (ISM) or
 intracluster medium (ICM), 
 and inflate a bubble composed of decelerated jet matter,
 which is often referred to as cocoon \citep[e.g.,][for review]{BBR84}.
 Since the cocoon is highly overpressured against the  
 ambient ISM/ICM  \citep{BC89}, bow shock is driven into the ambient  matter.
 As a result, the shocked ambient gas forms a thin shell around the
 cocoon and envelopes the whole system.
 The  thin shell structure persists until the cocoon pressure decreases and
 the pressure equilibrium is eventually achieved \citep{RHB01}.

 The dissipation of jet energy
 due to the interaction with the ambient matter is
 accompanied by  particle acceleration and non-thermal electrons
 are supplied to the cocoon from the jet. 
 It is well established that prominent radio emission
 is produced by the accelerated electrons 
 within the cocoon via synchrotron radiation.
Hence, the cocoon is often termed as radio lobe. 
 Inverse Compton (IC) emission by the same population of electrons 
 is sometimes observed at higher frequencies 
 such as X-ray \citep[e.g.,][]{HBC02, ITM02, CHH05, KS05}
 and, in some cases, even at gamma-rays \citep{Abd10}.

 As in the case of the lobe,
 shell is also expected to give rise to non-thermal emission,
 since 
 the bow shock offers a site of particle acceleration
\citep{FKY07,B08,BBP11,IKK11, KIK13}.
 Observationally, however, 
 detection of shell emission is rare
 and reported only at X-ray energies \citep{KVF03,CKH07,JHP08}.  
 Moreover, most of them are thermal origin,
 and only one source, 
 the most nearby radio galaxy Centaurus A,
 shows evidence for the non-thermal emission at X-ray \citep{CKH09}.
 Due to the lack of observations, the number of studies that focus
 on the shell emission is small and, therefore, its nature is poorly known.

 To tackle this issue,
 we have explored the evolution of
 the emission from both the lobe and shell simultaneously 
 in the previous studies \citep{IKK11, KIK13}.
 Under the assumption that the jet is active throughout the evolution,
 it is shown that the overall spectrum
 is strongly overwhelmed by the lobe emission at most of the frequencies.
 Hence, detection of the shell emission is likely to be hampered by
 the lobe emission and limited in a narrow range of frequencies.
 This provides a natural explanation on the reason
 why the detection of shell is rare.

 While the large contrast between the lobe and shell emission
 is expected to be maintained 
 as long as the jet continuously injects energy into the lobe,
 the situation may change drastically after the jet activity has ceased.
 It is known from the literature \citep[e.g.,][]{KB94,N10, MFB11} that 
 luminosity of the 
 lobe emission shows a rapid decrease  when the jet activity is stopped.
 This is because the radiative cooling leads to the 
 depletion of  non-thermal electrons,  
 since fresh electrons are no longer supplied from the jet.
 Therefore, it is often claimed 
 that  these  dead sources simply fade out 
 soon after the jet activity ceases. 
 However, 
 the above statement  may not be true when contribution from the shell emission
 is considered.
 Since the bow shock propagating into the ambient medium is active
 even after the jet activity has ceased,
 it continues to supply fresh electrons in the shell
 \citep{RB97, RHB02, PQM11}.
 As a result, emission from the shell can persist its luminosity
 and eventually dominates over that from the lobe.
 However, up to present, there are no quantitative studies that focus
 on this issue.

Regarding the evolutionarily track of radio sources, it 
is widely accepted that young compact radio sources,
i.e., Gigahertz Peaked Spectrum (GPS) and Compact Steep Spectrum (CSS) sources,
represent the young stage
\citep[$\sim 10^3-10^5~{\rm yr}$; e.g.,][]{O98}
 of typical well-extended classical radio galaxies
 that have estimated ages of
 $\sim 10^7-10^8~{\rm yr}$.
On the other hand, it is not clear whether ``all'' compact radio sources
can survive to the ages of typical radio galaxies.
This comes from observational fact that these compact sources 
represent a large fraction ($\sim 15-30\%$)  in
the flux-limited catalogue of the radios sources \citep[][]{FFD95},
which is much larger than the expected value
 ($\lesssim 0.1 \%$) from their youthness.
A simple and plausible explanation for this discrepancy is that 
 significant fraction of young compact radio sources
  stop their jet activity at an early stage of their evolution
  ($\lesssim 10^5~{\rm yr}$) 
 \citep{A00, MSK03, GTP05, OD10, OMD10,  KGL10}.
Then, large population
of compact dead radio sources in which jet injection from central engine
has ceased is expected to be hidden in the universe.
 Therefore, clarifying the evolution of dead radio sources 
 including the contribution of shell is essential for revealing
 the evolution of radio sources.

 Motivated by these backgrounds,
 we explore
 the emission from radio sources in which 
 jet activity has ceased at early stage of their evolution 
 in the present study.
 In particular, we focus on the evolution of the
 relative contribution of the lobe and shell emission.
 We show that the shell will be dominant at most of 
 the frequencies from radio up to gamma-ray soon
 after the jet is switched off and discuss the 
 possibility for detecting the emission.

 This paper is organized as follows.
 In Section \ref{model}, we introduce our dynamical model, which
 describes the evolution of the shell and cocoon, and 
 explain how the energy distribution of the electrons
 residing in these regions
 and the spectra  of the radiation they produce are evaluated.
 The obtained results are presented in Section \ref{results}.
 The summary and discussion of our results is given in Section \ref{sad}.
Throughout the paper, a $\Lambda$CDM cosmology with
 $H_0=71~{\rm km~s^{-1}~Mpc^{-1}}$, $\Omega_{\rm M}=0.27$ and 
$\Omega_{\Lambda}=0.73$ is adopted \citep{KDN09}.

\section{MODEL}
\label{model}


 Following the previous studies \citep{IKK11, KIK13}, we evaluate 
 non-thermal emissions from cocoon and shell based on a
 simple analytical model that describes their dynamical expansion.
 The model is identical to that employed in the previous studies
 during the early phase when the jet is active.
 The only difference 
 is that we properly take into account  
 the evolution after the
 jet injection has ceased. 
 In this section, we briefly review the employed model
 and explain the differences from the previous studies.

First, let us summarize the basic assumptions in our model.
 (1) Regarding the ambient mass density, $\rho_{\rm a}$,  
 a  power-law dependence on radius, $r$, with an index, $\alpha$,
 is assumed, $\rho_{\rm a}(r) ={\rho_0}(r/{ r_0})^{- \alpha}$.
 Here  ${r_0}$ is the reference radius and
 ${\rho_0}$ is the mass density at the radius.
 In the present study,
 we adopt
$\rho_0 = 0.1 m_p~{\rm cm^{-3}}, r_0 =  1~{\rm kpc}$ and $\alpha = 1.5$
based on the values inferred from the observation of   elliptical galaxies
\citep[e.g.,][]{MZ98, MB03, FMO04, FBP06}. 
 Here  $m_{p}$ is the proton mass and $\rho_{0.1} = \rho_0 / 0.1 m_p$. 
 The shape of the cocoon and shell are approximated as 
 a sphere and neglect its elongation along the jet direction merely 
 for simplicity.
 (2) We use the standard thin shell approximation \citep{OM88} and
 assume that most of the
 matter swept up by the bow shock is concentrated in
 a region of width , $\delta R$,
 behind the shock which is thin compared 
 with the radius of the shock, $R$.

Regarding the dynamics  of expansion, we consider two phases 
 depending on the age, $t$:
 (i) the early phase in which the jet injection is present  ($t\leq t_{\rm j}$) and
 (ii) the late phase in which the jet injection has ceased ($t>t_{\rm j}$).
Here  $t_{\rm j}$ denotes the duration of the jet injection.
As mentioned above, we use the same model used in
the previous studies \citep{IKK11, KIK13}
in the early phase ($\dot{R} \equiv dR/dt \propto t^{-(2-\alpha)/(5-\alpha)}$).
 After the energy injection from the jet ceases,  
 the cocoon will rapidly lose its energy due to adiabatic expansion
 and give away most of its energy into the shell within in a dynamical timescale $\sim R/\dot{R}$.
 Hence, after the transition time $\sim R/\dot{R}\sim t_{\rm j}$,
 cocoon pressure becomes dynamically unimportant, and the 
 energy of the shell  becomes dominant.
 Therefore, in the late phase, 
 expansion of the bow shock is expected to akin to
 Sedov-Taylor expansion  
 ($\dot{R} \propto t^{-(3-\alpha)/(5-\alpha)}$).
 Based on these considerations,
 we can approximately describe the two phases continuously by
 connecting the  the expansion velocity of the bow shock
 at $t=t_{\rm j}$ as 
\begin{eqnarray}
\label{V}
 \dot{R}(t) =
  \left\{ \begin{array}{ll}
      \dot{R}_{0} \left( \frac{t}{t_{\rm j}} \right)^{-(2-\alpha)/(5-\alpha)} &~~
         {\rm for}~~ 0<t \leq t_{\rm j} ,  \\
      \dot{R}_{0} \left( \frac{t}{t_{\rm j}} \right)^{-(3-\alpha)/(5-\alpha)} &~~
         {\rm for}~~ t_{\rm j} < t , \\
 \end{array} \right.
\end{eqnarray}
 where
\begin{eqnarray}
 \dot{R}_0 = C \left( \frac{L_{\rm j}}{\rho_0 r_0^{\alpha}} \right)^{1/(5-\alpha)}t_{\rm j}^{(2-\alpha)/(5-\alpha)} .
\end{eqnarray}
 Here
 $L_{\rm j}$ is the power of the jet and
 $C = 3/(5-\alpha)
 [ (3-\alpha)(5-\alpha)^3(\hat{\gamma_{\rm c}}-1) /
  \{4\pi ( 2\alpha^2 +  \alpha  - 18\hat{\gamma_{\rm c}}\alpha
    + 63\hat{\gamma_{\rm c}}-28 )  \} ]^{1/(5-\alpha)}$,
 where $\hat{\gamma}_{\rm c}$ is
 the specific heat ratio of the plasma inside the cocoon.
 From the above equation, the radius of the bow shock is determined
 as $R(t) = \int^{t}_0 \dot{R}(t') dt'$.
%
  In the present study, we assume
  $\hat{\gamma}_{\rm c} = 4/3$, since plasma
 within the cocoon is expected to be relativistic. 

Once the expansion velocity $\dot{R}$ is obtained,
the properties of the shell are determined
in the same manner described in the previous study \citep{IKK11}.
We employ the non-relativistic Rankine-Hugoniot
 condition in the strong shock limit~\citep{LL59} and evaluate
the density and pressure in the shell as
\begin{eqnarray}
\rho_{\rm s}(t)
             =  \frac{{\hat \gamma}_{\rm a} + 1}{{{\hat \gamma}_{\rm a} - 1}}
	          \rho_{\rm a}(R),
\end{eqnarray}
 and
\begin{eqnarray}
P_{\rm s}(t) = \frac{2}{{\hat{\gamma}_{\rm a} + 1}}
             \rho_{\rm a}(R) \dot{R}(t)^2   ,
\end{eqnarray}
 respectively,
 where ${\hat \gamma}_{\rm a}$ is the specific heat ratio
  of the ambient medium.
 In the present study, we adopt ${\hat \gamma}_{\rm a} = 5/3$
 since the temperature of the ambient medium is non-relativistic.
 The shell width, $\delta R$,
 in which most of the mass swept up by the bow shock is contained
 is evaluated from the conservation of mass
 $\rho_{\rm s} V_{\rm s} =\int_{0}^{R}4 \pi r^2 \rho_{\rm a}(r) dr$,
 where $V_{\rm s} = 4 \pi R^2 \delta R$ is the volume of the shell.
 Hence, the shell width is given by
$\delta R = ({\hat \gamma}_{\rm a} - 1)
  [({\hat \gamma}_{\rm a} + 1)(3-\alpha)]^{-1} R $.
%
%
 As a result,
the total internal energy of the shell, 
  $E_{\rm s} = P_{\rm s} V_{\rm s} / ({\hat{\gamma}_{\rm a}} - 1)$,
 scales linearly with time as
 $E_{\rm s} = f_1 L_{\rm j} t \propto t$ for $t\leq t_{\rm j}$ and gradually 
asymptotes to
 a constant value of $E_{\rm s} = f_2 L_{\rm j} t_{\rm j} \propto t^0$
 for $t \gg t_{\rm j}$.
 Such a behaviour is naturally expected, because
 energy is continuously injected into the system from the jet 
 with a constant rate
 when the jet is active, while there is no energy source after
  the jet activity ceases.
 Here
 $f_1$ and $f_2$ are the fractions of  energy injected by the 
  jet that are converted into the internal energy of the shell
 which are given 
 by $f_1 = 18 ({\hat{\gamma}_{\rm c}} - 1) (5-\alpha)
           ({\hat{\gamma}_{\rm a}} + 1)^{-2}
             [2\alpha^2 + (1-18{\hat{\gamma}_{\rm c}}) \alpha +
              63 {\hat{\gamma}_{\rm c}} - 28]^{-1} $ and
	     $f_2 = (3/2)^{3-\alpha} f_1$.
 For the values employed in the present study, (${\hat \gamma}_{\rm c} = 4/3$, 
  ${\hat \gamma}_{\rm a} = 5/3$ and $\alpha = 1.5$),  $f_1 \sim 0.11$ and $f_2 \sim 0.21$
 are obtained.
Hence, the ratio of the internal energy of the shell to
the total energy deposited in the system is
 roughly constant ($\sim 10-20 \%$) throughout the evolution.

 As described in the previous study
 \citep[see \S2.1 of][for detail]{IKK11},
 for $t \leq t_{\rm j}$,
 the pressure of the cocoon is directly obtained from the dynamical model
 and is given by 
\begin{eqnarray}
 P_{\rm c}(t) = f_{p} P_{\rm s}(t),
\end{eqnarray}
 where
 $f_p$ is the ratio of the pressure of the cocoon to that of shell
 and is given as
 $f_p = ({\hat \gamma}_{\rm a} + 1) (7 - 2\alpha) / [6(3-\alpha)]$.
 For typical numbers,
  ${\hat \gamma}_{\rm a} = 5/3$,
 $0 \leq \alpha \leq 2$, $f_p$ depends on $\alpha$
 only weakly and $f_p \sim 1$.
 This is expected, since the pressure of the shell and cocoon 
 should be roughly equal \citep[see e.g.,][]{HRB98}.
 Since pressure balance is also expected for $t > t_{\rm j}$,
 we assume that the ratio of the pressure is maintained 
 and determine $P_{\rm c}$ from the same equation also in
 the later phase.
 Regarding the evolution of 
 the radius of the cocoon, $R_{\rm c}$,
 we assume $R_{\rm c} = R$ for  $t \leq t_{\rm j}$ as in the previous studies.
 On the other hand,
 since the cocoon expands adiabatically,
 the cocoon radius satisfies
 $P_{\rm c}V_{\rm c}^{{\hat \gamma}_{\rm c}} = {\rm const}$,
  for   $t > t_{\rm j}$, where 
 $V_{\rm c} = 4\pi R_{\rm c}^3 / 3 $ is the volume of the cocoon.
 Hence, the evolution of the cocoon radius can be summarized as
\begin{eqnarray}
\label{Rc}
 R_{\rm c}(t) =
  \left\{ \begin{array}{ll}
     R(t) &~~
         {\rm for}~~ 0<t \leq t_{\rm j} ,  \\
      R(t_{\rm j}) \left(\frac{P_{\rm c}(t_{\rm j})}{P_{\rm c}(t)} \right)^{1/3{\hat \gamma}_{\rm c}}  &~~
         {\rm for}~~ t_{\rm j} < t . \\
 \end{array} \right.
\end{eqnarray} 
 As a result, cocoon expands slower than the shell
 in this phase and makes the region of the shocked shell wider.
 It is worth noting that similar behaviour is found in \citet{RB97}.
 The total
 internal energy deposited in the cocoon is evaluated
 from the obtained pressure and radius as
 $E_{\rm c} = P_{\rm c} V_{\rm c} / ({\hat{\gamma}_{\rm c}} - 1)$.
 For $t \leq t_{\rm j}$, $E_{\rm c}$ scales linearly with time as
 $E_{\rm c} = f_{\rm lobe,1} L_{\rm j} t$ as in the case of the shell,
 where 
 $f_{\rm lobe,1} = (5 - \alpha) (7 - 2 \alpha)
 [2 \alpha^2 + (1 - 18\hat{\gamma}_{\rm c})\alpha + 63 \hat{\gamma}_{\rm c} - 28]^{-1} $. 
 At later time ($t > t_{\rm j}$), the evolution of $E_{\rm c}$ asymptotes to
 $E_{\rm c} = f_{\rm lobe,2} L_{\rm j} t_{\rm j}
  (t/t_{\rm j})^{-6({\hat \gamma}_{\rm c} - 1)/
   [{\hat \gamma}_{\rm c}(5-\alpha)]}$, where
 $f_{\rm lobe,2} =
  (2/3)^{\alpha({\hat \gamma}_{\rm c} - 1)/{\hat \gamma}_{\rm c}}
   f_{\rm lobe,1}$.
 For the values employed in the present study, (${\hat \gamma}_{\rm c} = 4/3$, 
  ${\hat \gamma}_{\rm a} = 5/3$ and $\alpha = 1.5$),  $f_{\rm lobe1} \sim 0.54$ and $f_{\rm lobe,2} \sim 0.46$
 are obtained.

 In determining  energy distribution of
 non-thermal electrons, $N(\gamma_e, t)$, within the shell and lobe,
 we approximate each region as one-zone and
 solve the kinetic equation which takes into account
 the injection and cooling of particles
 given by
\begin{eqnarray}
 \label{kinetic}
  \frac{\partial N(\gamma_e, t)}{\partial t} =
 \frac{\partial}{\partial \gamma_e}
  [ \dot{\gamma}_{\rm cool}(\gamma_e, t) N(\gamma_e, t) ]
 + Q(\gamma_e, t) .
\end{eqnarray}
 Here $\gamma_e$, $\dot{\gamma}_{\rm cool} = - d\gamma_e/dt$,
 and $Q$ are the Lorentz factor, the cooling rate via adiabatic and radiative losses, and the injection rate of accelerated electrons, respectively.
 The injection rate and the cooling rate are determined from the
 dynamical model in the same manner as described in 
 the previous studies \citep{IKK11, KIK13}. \footnote{
 We note that
 the  temporal evolutions of
 $\dot{\gamma}_{\rm cool}$ and $Q$
 are taken into account in this paper and \citet{KIK13}, 
 whereas they were neglected in
 \citet{IKK11}.}


 Regarding the shell,
 we assume a continuous injection throughout the evolutions
 given by
\begin{eqnarray}
 \label{Q}
Q(\gamma_e, t) =
                K(t) \gamma_e^{-p}
         ~~{\rm for}~~  \gamma_{\rm min}\leq \gamma_e
                       \leq \gamma_{\rm max}(t) .
\end{eqnarray}
 Here
 $\gamma_{\rm min}$ and $\gamma_{\rm max}$ are
 the minimum and maximum Lorentz factors of
 the accelerated electrons, respectively.  
 In the present study, we employ $\gamma_{\rm min} = 1$ and $p = 2$.
 In determining the maximum Lorentz factor,
 we assume that the acceleration is limited by cooling.
As for the electron acceleration mechanism, we assume the
well-known diffusive shock acceleration in which the acceleration rate 
can be written as
  $\dot{\gamma}_{\rm accel} = 3 e B \dot{R}^2/(20 \xi  m_e c^3)$ \citep[e.g.,][]{D83}.
Here $B$ is  the magnetic field strength in the shell
 and $\xi$ is
 the so-called ``gyro-factor'' which can be expressed as
 the ratio of the energy
 in ordered magnetic fields to that in
 turbulent ones ($\xi = 1$ corresponds the Bohm limit).
 Hence, the maximum Lorentz factor
 is given by the value where
 cooling rate, $\dot{\gamma}_{\rm cool}(\gamma_e)$,  and 
 acceleration rate
 $\dot{\gamma}_{\rm accel}$ become equal.
 The normalisation factor, $K$, 
 is computed by assuming   
 that a fraction, $\epsilon_e$, of the
 shock-dissipated energy is carried by the 
 non-thermal electrons.
 In our calculation,
 we assume constant energy injection rate throughout the evolution,
 and,  for a given source age of $t = t_{\rm age}$, 
 the normalization factor at an arbitrary time 
 ($t~ (\leq t_{\rm age})$), is determined from the equation: 
 $\int^{\gamma_{\rm max}(t)}_{\gamma_{\rm min}} 
   (\gamma_e - 1) m_e c^2 Q(\gamma_e, t) d\gamma_e=
 \epsilon_e E_{\rm s}(t_{\rm age})/ t_{\rm age} $.
 We choose $\xi = 1$ and 
 and $\epsilon_e = 0.01$ as fiducial values
 for the parameters that characterize the acceleration efficiencies
 based on
 the observations of supernova remnants 
 \citep[e.g.,][]{DRB01, ESG01, BYU03, YYT04, SAH06, T08},
 since the properties of the shock are similar to that of the bow shock
 considered here.

 Considering the electron injection into the lobe,
 we assume continuous injection as in the case of the shell
 for $t \leq t_{\rm j}$.
 In contrast, 
 it is assumed that injection of the non-thermal electrons ceases
 for $t > t_{\rm j}$,
 since the jet can no longer supply matter in this phase.
 Hence,
 the energy distribution of the electrons injected into the lobe
 is given as 
\begin{eqnarray}
\label{Qc}
 Q_{\rm lobe}(\gamma_e, t) = 
  \left\{ \begin{array}{ll}
       K_{\rm lobe}(t) \gamma_e^{-p_{\rm lobe}} ~~~~~(0<t \leq t_{\rm j})   \\     
         ~~~{\rm for}~  \gamma_{\rm min, lobe}\leq \gamma_e   \leq \gamma_{\rm max, lobe},  \\
         0    ~~~~~
         (t_{\rm j} < t) , \\
 \end{array} \right.
\end{eqnarray}
 where 
 $\gamma_{\rm min, lobe}$ and $\gamma_{\rm max, lobe}$ are
 the minimum and maximum Lorentz factors, respectively.
 The imposed values for the power-law index 
 and minimum Lorentz factor are the same as those
 adopted for the shell
 ($p_{\rm lobe} = 2$ and $\gamma_{\rm min, lobe} = 1$).
 On the other hand, we use  as fixed value  $\gamma_{\rm max, lobe} = 10^4$
 for  the maximum Lorentz factor. 
 The employed value of  $\gamma_{\rm max, lobe}$ 
 is based on multi-wavelength observations of                     
 FRII radio galaxies which suggest electrons within the lobe do not have
 Lorentz factor well beyond
 $\sim 10^4$ \citep{SCH07, G09, YTI10}.
 As  shown in \S\ref{results},
 it is noted, however, 
 that the conclusion of the present study is insensitive to
 the assumed values of
 $p_{\rm lobe}$, $\gamma_{\rm min, lobe}$ and  $\gamma_{\rm max, lobe}$.
 In determining $K_{\rm lobe}$, 
 we assume  that a  fraction, $\epsilon_{e, {\rm lobe}}$,
 of the energy deposited in the lobe is carried by
 non-thermal electrons and the 
 energy injection rate of the non-thermal electrons
 are constant up to $t = t_{\rm j}$:
 $\int^{\gamma_{\rm max, lobe}}_{\gamma_{\rm min, lobe}} 
   (\gamma_e - 1) m_e c^2 Q(\gamma_e, t) d\gamma_e=
 \epsilon_{e, {\rm lobe}} E_{\rm c}(t_{\rm j})/ t_{\rm j} =
 f_{\rm lobe, 1} L_{\rm j}$.
 Although the value of 
 $\epsilon_{e, {\rm lobe}}$ is not constrained very well,
 we adopt
 $\epsilon_{e, {\rm lobe}} = 1$ in the present study
 which is often assumed in previous studies \citep[e.g.,][]{SBM08, OMS10, IKK11}.

 Regarding energy loss rate, $\dot{\gamma}_{\rm cool}$,
 we take into account  radiative and adiabatic (expansion) losses.
 The adiabatic cooling 
 is evaluated from the expansion of the system
 as
  $\dot{\gamma}_{\rm ad} =  (\dot{R}/R) \gamma_e$ for the shell
 and
 $\dot{\gamma}_{\rm ad} =  (\dot{R}_{\rm c}/{R_{\rm c}})\gamma_e$
 for the lobe, where $\dot{R}_{\rm c} = dR_{\rm c}/ dt$.
 As for the radiative cooling, we consider  
 the synchrotron radiation, $\dot{\gamma}_{\rm syn}$, and the IC scattering $\dot{\gamma}_{\rm IC}$.

 The typical magnetic field strengths in elliptical galaxies and clusters of
 galaxies are inferred to be around  few ${\rm \mu G}$ 
 \citep[e.g.,][]{MS96, VMM01, CKB01, CT02, SCK05}.
 Therefore,
 we  adopt  magnetic field strength of
 $B = 10{\mu {\rm G}}$ in the shell
 for the calculation of synchrotron emission, 
 since shock compression leads to amplification by a factor of $\sim 1-4$.
  On the other hand,
 magnetic field strength within the lobe, $B_{\rm c}$, is determined
 under the assumption that a fraction  
 $\epsilon_{\rm B}$, of the energy $E_{\rm c}$ is 
 carried by the magnetic fields. 
 In the present study we assume
 $\epsilon_{\rm B} = 0.1$, 
 a magnetic field strength  factor of a
 few below the equipartition value which is in the range typically observed
 in radio galaxies \citep[e.g.,][]{ITM02,KLE03, CBH04, CHH05, KS05}.

 Regarding the IC scattering, full Klein-Nishina (KN) cross section is
 taken into account \citep{BG70} and, as a source of seed photon fields, 
 we consider
 UV emission
 from the accretion disc, IR emission from the dusty torus, stellar emission 
 in NIR from the host galaxy, synchrotron emission from the radio lobe
 and cosmic microwave background (CMB).
 We modelled the spectra of 
 the photons from the disc,  torus and host galaxy
 with a black-body spectra peaking at frequencies 
 given by
 $\nu_{\rm UV} = 2.4 \times 10^{15}~{\rm Hz}$, 
 $\nu_{\rm IR} = 1.0 \times 10^{13}~{\rm Hz}$ and 
 $\nu_{\rm NIR} = 1.0 \times 10^{14}~{\rm Hz}$, respectively, 
 As for the luminosities of the emissions,
 we adopt $L_{\rm UV} = L_{\rm IR} = 10^{45}~{\rm erg~s^{-1}}$
 for the disc and torus emission \citep[e.g.,][]{E94,J06}, and 
 $L_{\rm NIR} = 10^{44}~{\rm erg~s^{-1}}$ for
 the host galaxy emission \citep[e.g.,][]{RPC05}.
 CMB is given by a black-body  that has peak frequency and 
 energy density given by  
$\nu_{\rm CMB} = 1.6 \times 10^{11}(1+z)~{\rm Hz}$
 and
$U_{\rm CMB} \approx 4.2 \times 10^{-13}(1+z)^4 
   ~{\rm  erg~cm^{-3}}$,
 respectively,
 where $z$ is the cosmological redshift of the source.
 The spectrum of the seed photon field 
 of the synchrotron emission from the lobe is
 obtained in our calculation self-consistently.
 The IC cooling is determined by the given spectra
 and source size as in the previous study \citep[for detail, see \S2.2 and \S4.1 of][]{IKK11}.

 The energy distribution of
 non-thermal electrons, $N(\gamma_e, t)$,
 is obtained by putting
 the evaluated injection rate, $Q(\gamma_e, t)$,
 and the cooling rate, $\dot{\gamma}_{\rm cool}(\gamma_e,t)$,
 in Equation~(\ref{kinetic}). 
 The spectra of the synchrotron and IC emission are 
 self-consistently calculated   
 from the obtained energy distribution \citep[for detail, see \S2.3 and \S4.1 of][]{IKK11}. 

  As shown in the previous studies, thermal emission is 
  also an important ingredient when considering the shell emission
  \citep[e.g.,][]{HRB98, KA99, RHB01, ZBR03, BBP11}.
  Therefore,
  we also calculate the thermal bremsstrahlung emission from the shell
  in the present study.
  Under the assumption that most of the shock dissipated
  energy is converted into that of the thermal electrons,
  the luminosity is estimated as
  $L_{\rm \nu, brem} = 2^5 \pi e^6/({3 m_e c^3})
  [2 \pi/(3 k_{\rm B} m_e)   ]^{1/2} n_e^2
  T^{-1/2} {\rm e}^{-h \nu / k_{\rm B} T} \bar{g}_{ff}$,
  where $k_{\rm B}$ is  the Boltzmann constant, and
  $n_e = \rho_{\rm s} / m_p$, and
  $T = P_{\rm s} / 2 n_e k_{\rm B}$ are
  the number density and
  temperature of the thermal electrons within the shell, respectively.
  Here $\bar{g}_{ff}$ is the Gaunt factor \citep{RL79}
  and is set to be unity merely for simplicity.

\section{RESULTS}
\label{results}


 In this section, we
 show the time 
 evolutions of the energy distribution of non-thermal electrons
 and the resulting emission.
 As a fiducial case we focus on sources with jet power of 
$L_{\rm j} = 10^{45}~{\rm erg~s^{-1}}$.
 Since our aim is to explore the evolution of short-lived sources in which
 the energy injection into the lobe have stopped in the early stage of
 their evolution, 
 we set the duration of energy injection as $t_{\rm j} = 10^{5}~{\rm yr}$.
The chosen value of the age is around the upper end of the estimated
age of the compact radio source  \citep[e.g.,][]{M03, PC03}.
 The overall linear size at $t = t_{\rm j}$ is 
 $2R(t_{\rm j}) \sim 0.86~
 {\rho}_{0.1}^{-2/7}
                  L_{45}^{2/7}
		  t_{\rm j,5}^{6/7}~
                   {\rm kpc} $,         
  where  $L_{45} = L_{\rm j}/10^{45}~{\rm erg~s^{-1}}$ and
  $t_{\rm j,5} = t_{\rm j} / 10^5~{\rm yr}$.

It is noted that, 
since  spherical symmetry is assumed,
the  source size and age correspondence
predicted in our model
show relatively large discrepancy from those 
inferred from the observations
 \citep[e.g.,][]{OCP99, TMP00, C02}.
The actual source geometry is elongated  along the jet axis,
and, for a given source age,
 their linear sizes along
the jet direction and transverse direction are larger
and smaller 
 by a factor $\sim {\rm a~few}$, respectively.
It is emphasized, however, that the discrepancy
does not have significant effect on the essential features of our results,
since large axial ratio is rarely observed.
Moreover, the axial ratio is reduced after 
the jet activity ceases ($t>t_{\rm j}$),
i.e.  the phase that 
we mainly focus on this study,
since the thrust from the jet which causes the elongation is no longer present.
Therefore,
we can consider that the results obtained in the present study at a given age
$t$ apply to the sources whose linear extension is larger than $R(t)$
within a factor of few.


\subsection{Evolution of Non-thermal Electrons}
\label{ENE}

\begin{figure*}[htbp]
\begin{center} 
\includegraphics[width=16cm,keepaspectratio]{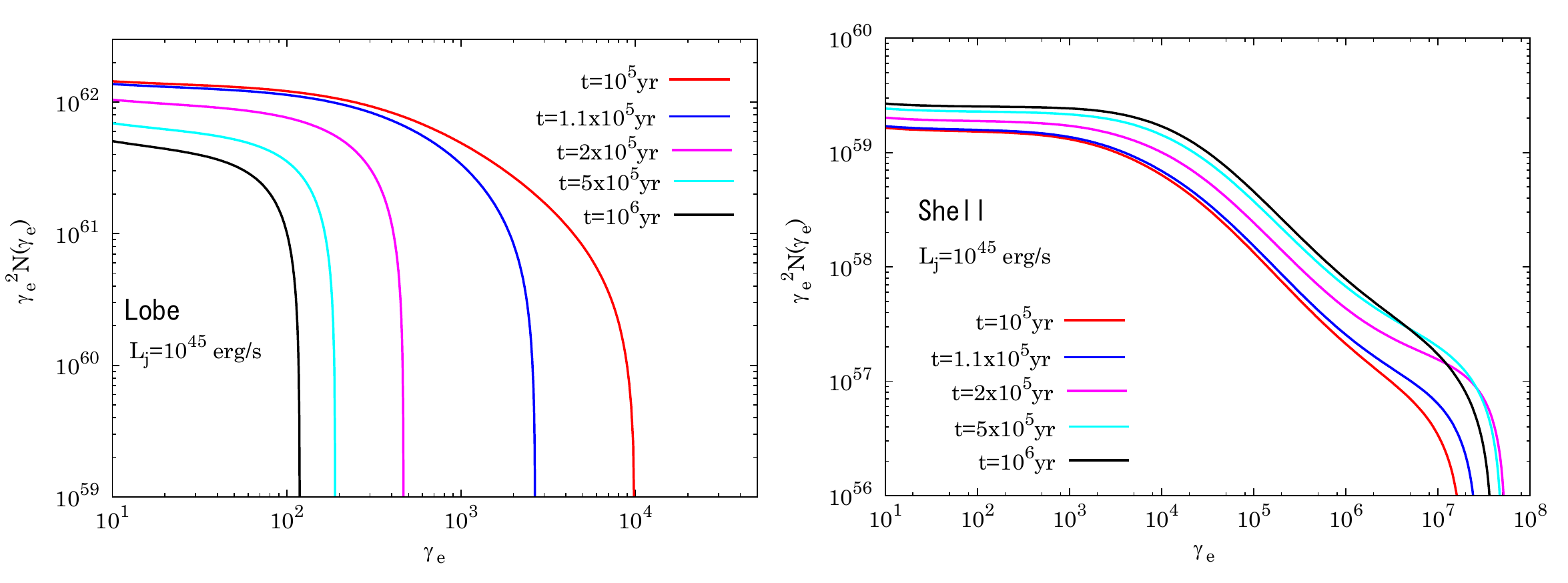}
\end{center}
\caption 
{
Energy distributions 
 of non-thermal electrons within the shell ({\it right panel})
 and lobe ({\it left panel}) 
 for source with jet power of  $L_{\rm j} = 10^{45}{\rm erg~s^{-1}}$ and
 injection duration of $t_{\rm j} = 10^5{\rm yr}$.
 The various lines display the cases for
 sources with ages of 
 $10^{5}$ (red line), 
 $1.1 \times 10^5$ (blue line),
 $2 \times 10^5$ (purple line),
 $5 \times 10^5$ (light blue line) and
 $10^6~{\rm yr}$ (black line).
} 
\label{Nevo}
\end{figure*}

 In Figure~\ref{Nevo}
 we display the energy distribution of non-thermal electrons
 within the lobe ({\it left panel}) and shell ({\it right panel})
 for different source ages 
($t=10^{5}$,  $1.1 \times 10^5$,
 $2 \times 10^{5}$,
 $5\times 10^{5}$ and $10^6~{\rm yr}$).
 When the jet is active ($t\leq t_{\rm j}$), the non-thermal electrons 
 are continuously supplied in the  shell and lobe.
 In this initial phase,
 the  energy distributions of electrons 
 within the lobe and shell 
 show a broken power-law shape. 
 The break is located at the energy 
 where the cooling time of the electrons
 becomes roughly equal to the dynamical time ($t\sim t_{\rm cool}(\gamma_e)$).
 Below the break energy,
 the electron energy distribution maintains the form of
 the injected energy spectrum which  is roughly given by
 $N(\gamma_e) \sim Q(\gamma_e)t \propto \gamma_e^{-2}$,
 since the cooling effect is negligible at these energies.
 Above the break energy, the energy distribution is modified by the cooling
 loss and  can be roughly approximated as
 $N(\gamma_e)\sim Q(\gamma_e)t_{\rm cool}(\gamma_e)$
 with a sharp cut-off at the injected maximum energy 
 ($\gamma_{\rm max}$ and $\gamma_{\rm max, lobe}$, 
  for the shell and lobe, respectively).
 While
 for the shell
 the dominant radiative cooling is provided by the IC scattering off
 the disc and torus photons,
 synchrotron cooling is slightly 
 larger than the IC cooling  for the lobe.
 Since the cooling is more rapid in the lobe, 
 the break energy is  lower than that of the shell
 \citep[see][for more detail]{IKK11}.

 After the jet activity has ceased ($t>t_{\rm j}$),
the evolution of electron energy distribution
differs largely between the shell and lobe.
 Regarding the shell,
 the evolution does not show large change from that
 at the early phase ($t\leq t_{\rm j}$).
 The electron energy distribution 
 can be roughly described by a broken power-law shape.
 The radiative loss is mainly dominated by the IC scattering 
 of  the  torus photons.
 Since the energy density of the photon decreases with
 the source size $U_{\rm ph} \propto R^{-2}$,
 the break energy gradually increases with the age. 
%
%

 On the other hand, 
 due to the cessation  of the electron injection ($Q(\gamma_e) = 0$),
 the evolution of
 the  energy distribution of the electrons in the lobe
 changes drastically from that  at the early phase ($t\leq t_{\rm j}$). 
%
%
%
 Since fresh electrons are no longer supplied,
 population of electrons with energy above the break energy
 determined at $t=t_{\rm j}$
 rapidly decreases because of the cooling loss.
 The cooling proceeds predominantly through synchrotron emission
 and the corresponding break energy is found at
\begin{eqnarray}
 \gamma_{\rm br, lobe} & \approx & 3 m_e c / (4 \sigma_T U_{\rm B, lobe} t) \nonumber \\
  & \sim & 4.8 \times 10^2 \epsilon_{\rm B, -1} \rho_{0.1}^{-6/7} L_{45}^{-1/7} t_{\rm j,5}^{4/7}, 
\end{eqnarray}
 where
 $U_{\rm B, lobe} = B_{\rm lobe}^2 / 8 \pi$,
 $\epsilon_{\rm B, -1} = \epsilon_{\rm B}/ 0.1$
 and $t_{\rm j, 5}= t_{\rm j} / 10^5{\rm yr}$. 
 The rapid depletion continues until $t \sim 2t_{\rm j}$ 
 at which  all the high energy electrons ($\gamma_e \gtrsim \gamma_{\rm br, lobe}$) are cooled.
 Thereafter,
 electrons gradually cool 
 mainly due to  adiabatic loss
 ($\gamma_e \propto R_{\rm c}^{-1} \propto t^{-3/7}$). 
 These features are clearly seen in Figure~\ref{Nevo}.

\subsection{Evolution of Broadband Emission}

  In Figure~\ref{Fluxevo}, we show the corresponding
  evolution of the total photon fluxes, $\nu f_{\nu}$,
  from the shell (solid lines) and lobe (dashed lines)
  for a source located at the redshift of $z=1$.
  Together with the total photon fluxes,
  the contributions from the synchrotron emission, 
  and IC scattering of various seed photons
 are displayed  in Figure~\ref{Flux}
  for a source with $t = 10^5~{\rm yr}$ 
  and $10^6~{\rm yr}$.
  As mentioned in the previous section, contribution from 
  the thermal bremsstrahlung
  emission is also considered for the shell emission.
  Attenuations 
  due to synchrotron self-absorption (SSA)
  and pair production due to photon-photon interaction
  are also taken into account in the spectra.
  The absorption coefficient for SSA
  within the lobe, $\alpha_{\rm lobe, \nu}$,
  and shell $\alpha_{\rm s, \nu}$ are determined
  independently from the 
  magnetic field strength and electron energy distribution 
  within these regions following \citet{RL79}.
  The spectra of the shell and lobe below
   the  frequency  which satisfy
  $\alpha_{\rm s, \nu} R = 1$ and
  $\alpha_{\rm lobe,\nu} R_{\rm c} = 1$ (optically thick), respectively,
  are described as $f_{\nu} \propto \nu^{5/2}$.
  Note that the absorption in the lobe is much 
  stronger than that of the shell since 
  the number of non-thermal electrons
  and magnetic field strength within the lobe
  is larger than those of the shell.
  Gamma-rays above energy $\gtrsim 100~{\rm GeV}$ are subject
  to pair production due to photon-photon interaction 
  with optical-infrared background radiation.
  Here we used the model of \citet{FRV08} for the background radiation
  and evaluated the attenuated photon fluxes shown in
  Figures~\ref{Fluxevo} and \ref{Flux}.
  We also plot the unattenuated gamma-ray spectra in Figure~\ref{Flux}
  (dotted line).

\subsubsection{Lobe  emission}

  Regarding lobe emission, 
  synchrotron and IC components 
  produce prominent emission
  extending from radio up to $h \nu \sim 10~{\rm GeV}$
  when the jet is active ($t\leq t_{\rm j}$).
  Their overall  luminosities are roughly comparable and 
  proportional to
  energy injection rate into the non-thermal electrons
  ($\nu L_{\nu} \propto \epsilon_{e, {\rm lobe}}L_{\rm j}$).
%
  This is simply because the energy injected into the non-thermal electrons
  $ \epsilon_{e, {\rm lobe}}L_{\rm j}$
  is immediately converted into the radiation.

  After the jet activity ceases ($t>t_{\rm j}$),
 due to the absence of particle injection,
 radiative cooling leads to
 rapid depletion of high energy electrons
 $\gamma_e > \gamma_{\rm br, lobe}$ within the lobe.
 As mentioned earlier, this rapid decrease 
 continues until all the electrons above the break energy 
 radiate away most of their energy 
 $t\sim 2 t_{\rm j}$
  (Figure.~\ref{Nevo}).
 As a result, 
 the high energy cut-off frequency rapidly shifts to the 
 lower energies and
 the overall emission fades out
 quite rapidly,
 as  seen in  Figure~\ref{Fluxevo}.   
%
 It is worth noting that this passive feature of the lobe emission
 is consistent with  
 the previous theoretical studies \citep{KB94,N10, MFB11}
 and provides a natural explanation
 for the reason  why dying radio sources, so-called faders,
 are rarely found in the observations \citep{KMT05, KMT06,OMD10}.
 After this passive phase ($t \gtrsim 2 t_{\rm j}$),
 although relatively slow, 
 the emission continues to fade 
 predominantly through adiabatic cooling.
  It is stressed
  that the evolution at the fading phase ($t> t_{\rm j}$) is
  insensitive to the assumed
  value of $\gamma_{\rm max,lobe}$ 
  or the assumptions on the seed photon fields,
  since  $\gamma_{\rm br, lobe}$ does not depend on these 
  assumptions.

\subsubsection{Shell  emission}

  Regarding emission from the shell, as in the case of 
  the lobe,
  the overall luminosity of the non-thermal emission
  is  proportional  to the jet power
  and the energy fraction of the non-thermal electrons
  ($\nu L_{\nu} \propto \epsilon_e L_{\rm j}$)
  when the jet is active $t \leq t_{\rm j}$.
%
%
  However,
  much fainter emission is produced in the shell,
  since smaller energy fraction is assumed 
  ($\epsilon_e \ll \epsilon_{e, {\rm lobe}}$).
  Unlike in the case of the lobe,
  IC emission is brighter than the synchrotron emission,
  since the energy density of magnetic field, $U_{\rm B}=B^2/8\pi$
  is much smaller than that of the seed photons.
  ($U_{\rm B}\ll U_{\rm ph}$).
   Since the maximum energy of the electron is larger 
  than that of the lobe ($\gamma_{\rm max} \gg \gamma_{\rm max, lobe}$), 
  the overall spectrum extends up to higher frequencies.
 The high energy cut-off is observed at $h \nu \sim 1-10~{\rm TeV}$,
 and at this high energy range the spectrum is modified by the
 attenuation due to pair production (Figure~\ref{Flux}).

\begin{figure*}[htbp]
\begin{center} 
\includegraphics[width=15cm,keepaspectratio]{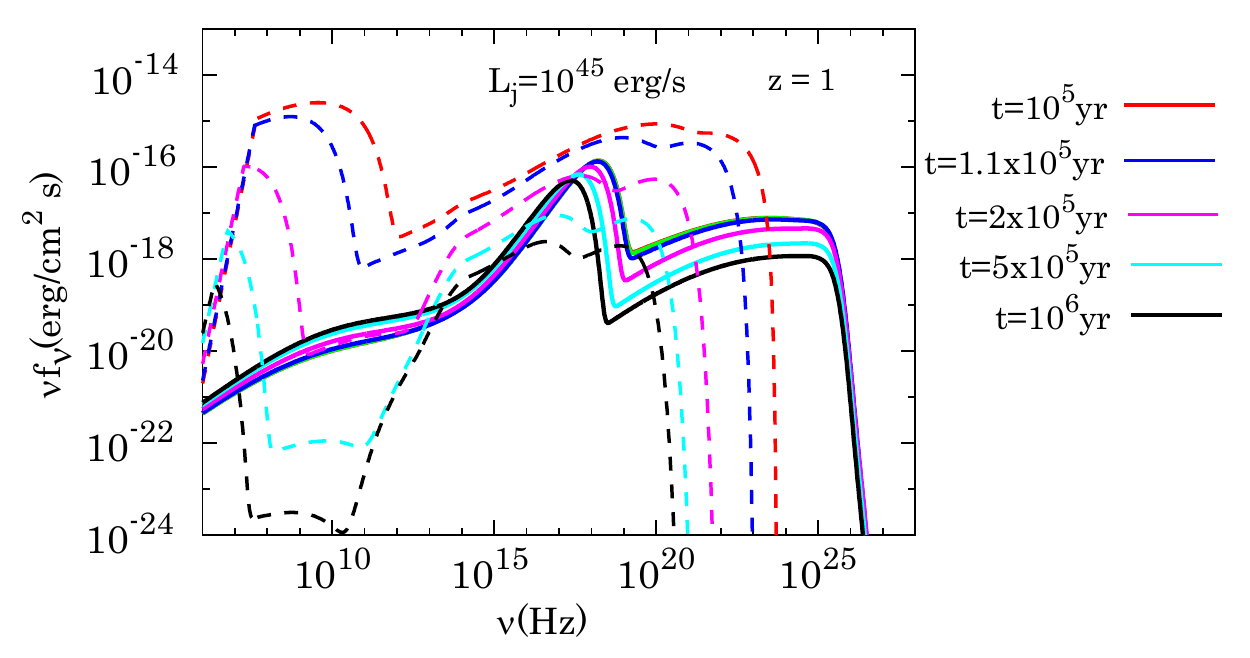}
\end{center} 
\caption 
{ Broadband spectrum  of 
 source with a jet power of
 $L_{\rm j} = 10^{45}{\rm erg~s^{-1}}$ that
 located at the redshift of $z=1$, and
 an injection time $t_{\rm j} = 10^5~{\rm yr}$.
 The dashed lines and solid lines display the
 emission produced within the lobe and shell, respectively, 
 for sources with age of
 $10^5~{\rm yr}$ (red line),     
 $1.1 \times 10^5~{\rm yr}$ (blue line),          
 $2 \times 10^5~{\rm yr}$ (purple line),          
 $5 \times 10^5~{\rm yr}$ (light blue line) and     
 $10^6~{\rm yr}$ (black).     
}
\label{Fluxevo}
\end{figure*}

 As mentioned in \S\ref{ENE}, the evolution of non-thermal electrons
 at the later phase ($t> t_{\rm j}$) 
 does not show large change from the initial phase  ($t\leq t_{\rm j}$).
%
 The overall luminosity  continues to be dominated by the IC emission.
 As in the case of the lobe, the emission from the 
 shell begins to fade in this phase.
 On the other hand, however,  
 their cut-off frequency $h \nu \sim 1-10~{\rm TeV}$
 remains nearly unchanged and the
 decrease in the luminosity proceeds much more slowly 
(Figure~\ref{Fluxevo}).
As in the initial phase,
the overall luminosity 
is proportional to the energy injection rate into the non-thermal electrons
($\nu L_{\nu} \propto \epsilon_e L_{\rm j}t_{\rm j} / t)$.
Hence, after the jet injection has ceased
the luminosity gradually decreases with age as 
$\nu L{\nu} \propto t^{-1}$, if the energy fraction of 
electrons $\epsilon_e$ doe not vary with time. 
 This large difference in the behaviour of luminosity
 evolution between the lobe and shell 
 is mainly due to the fact that
 the fresh 
 electrons that are accelerated at the bow shock
 are continuously supplied into the shell even after the jet cessation.

 In contrast with the non-thermal emission,
 the thermal bremsstrahlung emission from the shell
 has a relatively weak dependence on the jet power and 
 has a strong dependence on the ambient medium.
 The emission  produces a bump in the spectra at X-ray frequencies
 ($h \nu \sim 1-100~{\rm keV}$)
 with peak frequency 
 given as $h\nu \sim k_{\rm B} T \propto \dot{R}(t)^2$ 
 and dominates over the non-thermal emission 
 as  seen in the Figures.~\ref{Fluxevo} and \ref{Flux}.

\subsubsection{Comparison between lobe and shell}

 Finally, let us compare the 
 emission from the lobe and shell.
In the initial phase when the jet is active  ($t \leq t_{\rm j}$),
the result does not vary from those obtained
in the previous studies \citep{IKK11, KIK13}, since the
same set up is used for the  calculation.
The  lobe emission dominates over the shell emission
 up to their cut-off frequency  ($h \nu \sim 10~{\rm GeV}$).
 At higher energies, however, 
 the shell emission becomes dominant up to  $h \nu \sim 1-10~{\rm TeV}$.
%

\begin{figure*}[htbp]
\begin{center} 
\includegraphics[width=13cm,keepaspectratio]{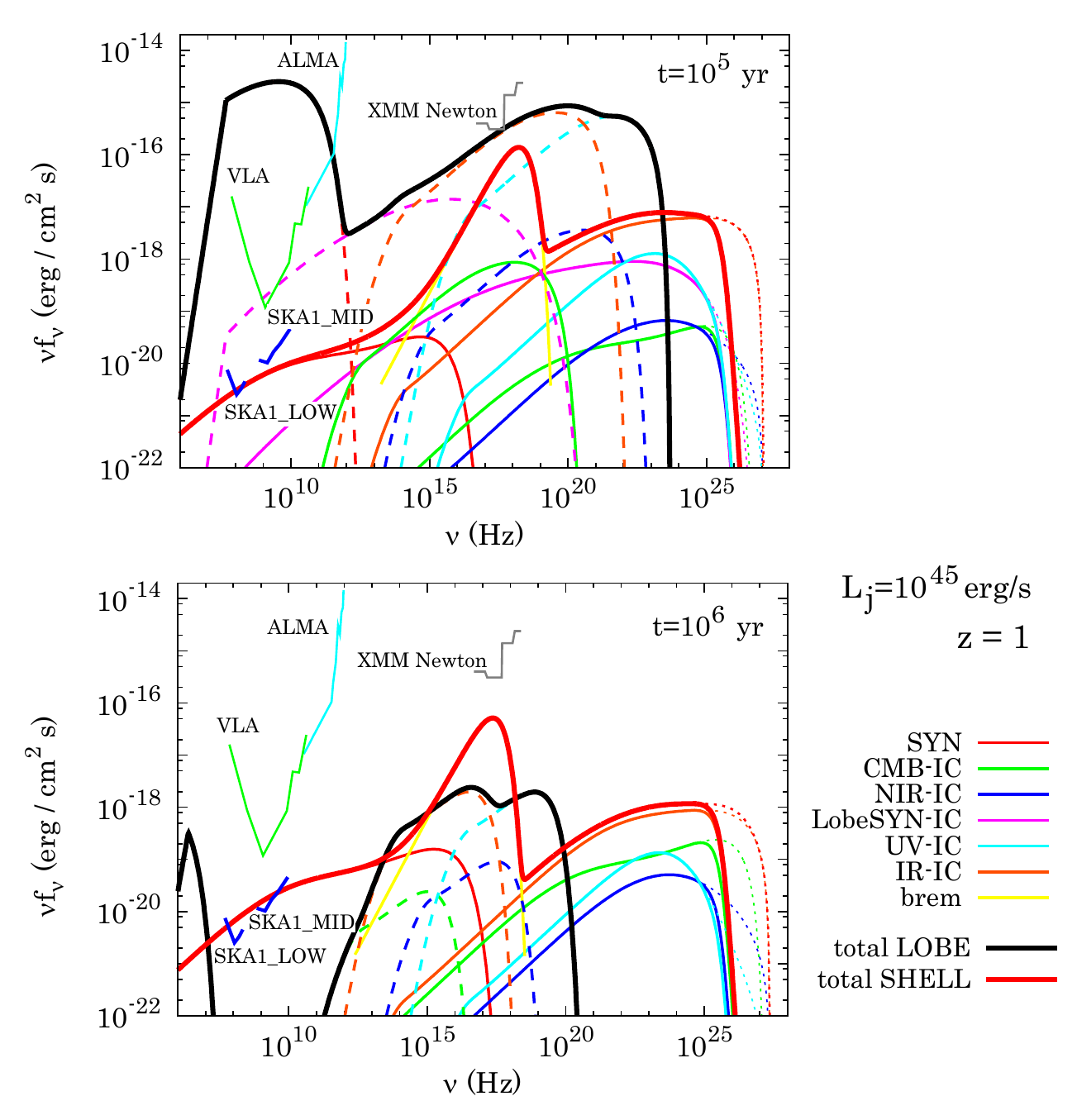}
\end{center} 
\caption 
{
 Broadband spectrum  of source with a 
 jet power of
 $L_{\rm j} = 10^{45}{\rm erg~s^{-1}}$ that is 
 located at the redshift of $z=1$.
 The top and bottom panels display the spectra for  sources with 
 ages of $10^5$ and $10^6~{\rm yr}$, respectively. 
 The thick black and thick red lines show the total flux from 
 the shell and lobe, respectively.
 The various thin dashed and thin solid lines show 
 the
 contributions to the total emission produced in the lobe
 and shell, respectively,
 from the synchrotron emission (red line)
 and the IC scattering of UV disk photons (light blue line),
 IR torus photons (orange line), 
 NIR host galaxy photons (blue line), 
 CMB photons (green line) and
 lobe synchrotron photons (purple line).
 The contribution from the thermal bremsstrahlung
 emission  is also shown for the shell (yellow line).
 The dotted lines show the case when the attenuation due to  
 pair production is neglected.
 Also shown are the sensitivities of the  XMM-Newton, 
ALMA, Jansky VLA and SKA1 (SKA1\_MID and SKA1\_LOW).
The assumed integration times for XMM-Newton,  Jansky VLA and ALMA
are 100 ks, 
4 and 10 hours, respectively.
The sensitivity of ALMA is calculated by using the 
 ALMA Sensitivity Calculator 
(https://almascience.eso.org/proposing/sensitivity-calculator).
As for SKA, we display 3$\sigma$ detection limit for an integration time of
10 hours which is calculated based on the Appendix of \citet[][]{PS14}.
}
\label{Flux}
\end{figure*}

 On the other hand,
 soon after the cessation of the jet activity ($t > t_{\rm j}$),
 emission from the shell becomes dominant 
 at most of the frequencies 
 due to the rapid decrease of the lobe luminosity.  
 For example,  the emission from the shell 
 overwhelms that of the lobe at frequencies $\gtrsim 10^{7}~{\rm Hz}$
 for sources with age of $t \gtrsim 10^{6}~{\rm yr}$
 (see Figure~\ref{Flux}).
 As a result, dead radio sources 
 eventually become dominated by the shell emission even at the
 radio frequencies.
This shell-dominated phase is expected to continue up to the
age where the 
expansion velocity becomes subsonic,
since particle acceleration is unlikely to take place
thereafter.
As discussed in next section, the corresponding age
can be as large as $\gtrsim 10^7~{\rm yr}$.
Until then, luminosity of the shell gradually decreases 
as $\propto t^{-1}$.
 It is noted that the candidate of dying radio sources
 found in the observations \citep{KMT05, KMT06,OMD10} 
 should be in passive phase ($t_{\rm j} < t < 2t_{\rm j}$)
 which means that the  lobe emission has started to fade, but 
 still dominates the emission at radio wavelengths.

\section{SUMMARY and DISCUSSIONS}\label{sad}

 In the present study,
 we have explored the evolution of the emission 
 from the lobe and shell
 of  radio sources which have ceased their jet activity
 in early stage of their evolution  ($t_{\rm j} \lesssim 10^5~{\rm yr}$).
 It is shown that the lobe emission
 rapidly fades after the jet injection has been stopped
 ($t > t_{\rm j}$) due to the absence of 
 particle supply from the jet.
 On the other hand,
 shell emission  only shows gradual decay
 even after the jet activity has stopped
 since supply of electrons from the bow shock is maintained.
 Therefore,  although faint,
 emission from the dying radio sources
 eventually becomes shell dominated
 at wide range of frequencies from radio up to gamma-ray.

Our results suggest that shell emission is essential for studying 
 the nature  of dead radio sources. 
 Shell emission may be detectable for  
 exceptionally high power sources
 \citep[$L_{\rm j}\gtrsim 10^{46}~{\rm erg~s^{-1}}$; e.g.,][]{IKK08} that 
 are located nearby ($z \lesssim 0.2$).
 On the other hand, 
 for typical sources ($L_{\rm j}\sim 10^{45}~{\rm erg~s^{-1}}$)
  located at redshift of $z \sim 1$,
 \citep[a typical value for compact radio sources; e.g.,][]{O98},
 the emission is below the thresholds of the current instruments
 at any frequencies (see Figure.~\ref{Flux}).
 This provides a natural explanation why such a shell emission
 has not been reported  so far.
 Regarding the future missions, 
 SKA telescope is capable of detecting the emission in the radio band.
As shown in Figure.~\ref{Flux}, the detection is marginal for 
SKA phase 1 (the emission slightly exceeds 3$\sigma$ detection limit for an
integration time of 10 hours with conservative values of $\epsilon_e=0.01$ and
$L_{\rm j}=10^{45}~{\rm erg~s^{-1}}$).
It is noted that the emission is roughly linearly proportional to 
the acceleration efficiency $\epsilon_e$ and jet power $L_{\rm j}$.
Since these values inevitably have dispersion from source to source,
we can expect a certain fraction of sources to be 
well above the detection limit.
Moreover, in the phase 2 (SKA2), the sensitivity is expected improve by
an order of magnitude.
Hence, SKA will be a powerful tool to reveal the 
population of dead radio sources which are dominated by the shell emission.

 In order to confirm that the emission comes from the shell rather than lobe,
 spatial angular resolution of the observation is important. 
 If the emission is indeed produced in the shell,
 we expect to see limb brightening features \citep{HRB98, BBP11}.
 The emission will appear as an extended source with 
 a total angular size given as
 $\sim 2 (2 R/5~{\rm kpc})(D_{\rm A}/1~{\rm Gpc})^{-1}~{\rm arcsec}$,
where $D_{\rm A}$ is the angular diameter ($D_{\rm A}\sim 1.6~{\rm Gpc}$ for $z=1$).
Hence, high spacial resolution ($\sim 0.1~~{\rm arcsec}$)
provided by SKA can detect such a surface brightness distribution.
%
%

 The shell emission dominated phase is expected 
 to persist until the expansion velocity $\dot{R}(t)$ becomes sub-sonic,
 since, thereafter, strong shock will no longer be present.
 The sound speed in ambient gas is given as 
 $c_s = [{\hat \gamma}_{\rm a}  k_{\rm B} T_{\rm a}/ (\mu m_p)]^{1/2}$,
 where $T_{\rm a}$ and $\mu = 0.6$ are 
 the temperature and mean molecular weight of the ambient gas, respectively.
 Then the age when the source becomes subsonic ($\dot{R}/c_s = 1$)
 can be estimated as 
$t_s \approx 8.4 \times 10^6 (L_{45}/\rho_{0.1})^{2/3} t_{\rm j,5}^{4/3} T_{\rm keV}^{-7/6}~{\rm yr}$, 
where
 $T_{\rm keV} = k_{\rm B}T_{\rm a} / 1~{\rm keV}$.
 From the above equation, it 
 is confirmed that the duration of shell dominated phase is
 much longer than that of the lobe dominated phase
 ($t_s - t_{\rm j} \gg t_{\rm j}$).
 Therefore, if large fraction of young radio sources die 
 before becoming classical extended radio
 galaxies ($t \sim 10^7 - 10^8~{\rm yr}$) as
 indicated in previous studies \citep{A00, MSK03, GTP05, OD10, OMD10,  KGL10},
 our results predict
 the existence of a 
 large population of
 undetected shell dominated sources residing in the universe.
 We expect that future survey by SKA will 
 enable us to perform systematic study of these sources
 and quantify the fraction of dying sources through the detection
 of their shell emission.
 These observations will provide an important constraint
 on the evolution of radio sources.

Lastly, let us comment on the radio sources
which stop their jet activity
after becoming  classical extended radio lobes \citep{FR74}.
Although we focused on the sources which die at and early stage 
of their evolution ($\lesssim 10^{5}~{\rm yr}$),
qualitatively similar results are expected  for such sources.
As shown in the present study, their lobe emissions fade rapidly
and eventually become shell dominated.
Actually, candidates of dead classical radio galaxies that are 
in the lobe fadying phase  
have been found in the observations \citep[e.g.,][]{MPM11,HJE15}.
Therefore, we also expect to find the scaled up version
of the compact shell dominated sources considered in the present study.
Such sources are also very interesting
 and detectable by the SKA.
Provided that the observed numbers of compact ($t\lesssim 10^5~{\rm yr}$)
 and well extended radio source ($t\sim 10^7-10^8~{\rm yr}$) 
are comparable \citep{FFD95},
the detection rate of such sources are 
 expected to be roughly comparable to
those end their jet activity
at an early stage of their evolution ($\lesssim 10^{5}~{\rm yr}$).
%
%





\acknowledgments 

We are grateful to the anonymous referee for constructive
comments which improved the clarity of the paper. 
N.K. acknowledges the financial support of Grant-in-Aid
for Young Scientists (B:25800099).
This work was partly supported by the Grant-in-Aid for
Young Scientists (B: 26800159) from
the Ministry of Education, Culture, Sports, Science and Technology, Japan.
Part of this work was done with the contribution of the Italian Ministry of Foreign Affairs and Research for the collaboration project between Italy and Japan.
%




\begin{thebibliography}{}



\bibitem[Abdo et al.(2010)]{Abd10} Abdo, A.~A., Ackermann,
M., Ajello, M., et al.\ 2010, Science, 328, 725


\bibitem[Alexander(2000)]{A00} Alexander, P.\ 2000, \mnras, 
319, 8 




\bibitem[Bamba et al.(2003)]{BYU03} Bamba, A., Yamazaki, R., 
Ueno, M., \& Koyama, K.\ 2003, \apj, 589, 827 




\bibitem[Begelman \& Cioffi(1989)]{BC89} Begelman, M. C.,
    \& Cioffi, D. F.  1989,  \apj, 345, 21


\bibitem[Begelman et al.(1984)]{BBR84} Begelman, M.~C., 
Blandford, R.~D., \& Rees, M.~J.\ 1984, Reviews of Modern Physics, 56, 255 

\bibitem[Berezhko(2008)]{B08} Berezhko, E.~G.\ 2008, \apjl, 
684, L69


\bibitem[Blumenthal 
\& Gould(1970)]{BG70} Blumenthal, G.~R., \& Gould, R.~J.\ 1970, Reviews of Modern Physics, 42, 237 


\bibitem[Bordas et al.(2011)]{BBP11} Bordas, P., Bosch-Ramon, 
V., \& Perucho, M.\ 2011, \mnras, 8 
 




\bibitem[Carilli et al.(1988)]{CPD88} Carilli, C.~L., Perley, 
R.~A., \& Dreher, J.~H.\ 1988, \apjl, 334, L73 

\bibitem[Carilli et al.(1994)]{CPH94} Carilli, C.~L., Perley, 
R.~A., \& Harris, D.~E.\ 1994, \mnras, 270, 173 



\bibitem[Carilli  \& Taylor(2002)]{CT02} 	
 Carilli, C. L., \& Taylor, G. B. 2002, ARA\&A, 40, 319






\bibitem[Clarke et al.(2001)]{CKB01} Clarke, T.~E., Kronberg, 
P.~P., \& B\"ohringer, H.\ 2001, \apjl, 547, L111 


\bibitem[Conway(2002)]{C02} Conway, J.~E.\ 2002, NewAR, 46, 263 


\bibitem[Croston et al.(2004)]{CBH04} Croston, J.~H., 
Birkinshaw, M., Hardcastle, M.~J., 
\& Worrall, D.~M.\ 2004, \mnras, 353, 879 



\bibitem[Croston et al.(2005)]{CHH05} Croston, J. L.,
 Hardcastle, M. H., Harris, D. E., Besole, E., Birkinshaw, M.,
 \& Worrall, D. M. 2005, \apj, 626, 733


\bibitem[Croston et al.(2007)]{CKH07} Croston, J.~H., Kraft, 
R.~P., \& Hardcastle, M.~J.\ 2007, \apj, 660, 191 



\bibitem[Croston et al.(2009)]{CKH09} Croston, J.~H., et al.\ 
2009, \mnras, 395, 1999 


\bibitem[Drury(1983)]{D83} Drury, L.~O.\ 1983, Reports on 
Progress in Physics, 46, 973 

\bibitem[Dyer et al.(2001)]{DRB01} Dyer, K.~K., Reynolds, 
S.~P., Borkowski, K.~J., Allen, G.~E., \& Petre, R.\ 2001, \apj, 551, 439 




\bibitem[Ellison et al.(2001)]{ESG01} Ellison, D.~C., Slane, 
P., \& Gaensler, B.~M.\ 2001, \apj, 563, 191 




\bibitem[Elvis et al.(1994)]{E94} Elvis, M., et al.\ 1994, 
\apjs, 95, 1 


\bibitem[Fanaroff 
\& Riley(1974)]{FR74} Fanaroff, B.~L., \& Riley, J.~M.\ 1974, \mnras, 167, 31P 


\bibitem[Fanti et 
al.(1995)]{FFD95} Fanti, C., Fanti, R., Dallacasa, D., et al.\ 1995, \aap, 302, 317 



\bibitem[Franceschini et al.(2008)]{FRV08} Franceschini, A., Rodighiero, G., \& Vaccari, M.\ 2008, \aap, 487, 837 


\bibitem[Fujita et al.(2007)]{FKY07}
Fujita, Y., Kohri, K., Yamazaki, R., \& Kino, M. 2007, \apj, 663, L61



\bibitem[Fukazawa et al.(2006)]{FBP06} Fukazawa, Y., 
Botoya-Nonesa, J.~G., Pu, J., Ohto, A., \& Kawano, N.\ 2006, \apj, 636, 698 


\bibitem[Fukazawa,  Makishima,  \& Ohashi(2004)]{FMO04} 
Fukazawa, Y., Makishima, K., \& Ohashi, T. 2004, PASJ, 56, 965






\bibitem[Godfrey et al.(2009)]{G09} Godfrey, L.~E.~H., et 
al.\ 2009, \apj, 695, 707 


\bibitem[Gugliucci et al.(2005)]{GTP05} Gugliucci, N.~E., 
Taylor, G.~B., Peck, A.~B., \& Giroletti, M.\ 2005, \apj, 622, 136 


\bibitem[Hardcastle et al.(2002)]{HBC02} Hardcastle, M.~J
., Birkinshaw, M., Cameron, R.~A., et al.\ 2002, \apj, 581, 948


\bibitem[Heinz et al.(1998)]{HRB98} Heinz, S., Reynolds, 
C.~S., \& Begelman, M.~C.\ 1998, \apj, 501, 126 

\bibitem[Hurley-Walker et al.(2015)]{HJE15} Hurley-Walker, 
N., Johnston-Hollitt, M., Ekers, R., et al.\ 2015, \mnras, 447, 2468 


\bibitem[Isobe et al.(2002)]{ITM02} 
Isobe, N., Tashiro, M., Makishima, K., Iyomoto, N., Suzuki, M., Murakami, M. M., Mori, M., \& Abe, K. 2002, \apj, 580, L111

\bibitem[Ito et al.(2008)]{IKK08} Ito, H., Kino, M., 
Kawakatu, N., Isobe, N., \& Yamada, S.\ 2008, \apj, 685, 828 				


\bibitem[Ito et al.(2011)]{IKK11} Ito, H., Kino, M., 
Kawakatu, N., \& Yamada, S.\ 2011, \apj, 730, 120 


\bibitem[Jetha et al.(2008)]{JHP08} Jetha, N.~N., Hardcastle, 
M.~J., Ponman, T.~J., \& Sakelliou, I.\ 2008, \mnras, 391, 1052 


\bibitem[Jiang et al.(2006)]{J06} Jiang, L., et al.\ 2006, 
\aj, 132, 2127 





\bibitem[Kaiser \& Alexander(1999)]{KA99} 	
Kaiser, C. R., \& Alexander, P. 1999, \mnras, 305, 707


\bibitem[Kataoka et al.(2003)]{KLE03} Kataoka, J., Leahy, J.~P., Edwards, P.~G., et al.\ 2003, \aap, 410, 833 



\bibitem[Kataoka \& Stawarz(2005)]{KS05} Kataoka, J.,
 \& Stawarz, L. 2005, \apj, 622, 797


\bibitem[Kino et al.(2013)]{KIK13} Kino, M., Ito, H., 
Kawakatu, N., \& Orienti, M.\ 2013, \apj, 764, 134 

\bibitem[Komatsu et al.(2009)]{KDN09} Komatsu, E., Dunkley, 
J., Nolta, M.~R., et al.\ 2009, \apjs, 180, 330 


\bibitem[Komissarov \& Gubanov(1994)]{KB94} Komissarov, S.~S., \& Gubanov, A.~G.\ 1994, \aap, 285, 27 


\bibitem[Kraft et al.(2003)]{KVF03} Kraft, R.~P., 
V{\'a}zquez, S.~E., Forman, W.~R., et al.\ 2003, \apj, 592, 129


\bibitem[Kunert-Bajraszewska et al.(2010)]{KGL10}
Kunert-Bajraszewska, M., Gawro{\'n}ski, M.~P., Labiano, A., 
\& Siemiginowska, A.\ 2010, \mnras, 408, 2261 


\bibitem[Kunert-Bajraszewska et 
al.(2005)]{KMT05} Kunert-Bajraszewska, M., Marecki, A., Thomasson, P., \& Spencer, R.~E.\ 2005, \aap, 440, 93 



\bibitem[Kunert-Bajraszewska et 
al.(2006)]{KMT06} Kunert-Bajraszewska, M., Marecki, A., \& Thomasson, P.\ 2006, \aap, 450, 945 



\bibitem[Landau \&  Lifshitz(1959)]{LL59} 
Landau, L., \&  Lifshitz, F. M. 1959, Fluid Mechanics (London: Pergamon)


\bibitem[Marecki et al.(2003)]{MSK03} Marecki, A., Spencer, 
R.~E., \& Kunert, M.\ 2003, PASA, 20, 46 

\bibitem[Mathews 
\& Brighenti(2003)]{MB03} Mathews, W.~G., \& Brighenti, F.\ 2003, \araa, 41, 191 



\bibitem[Mocz et al.(2011)]{MFB11} Mocz, P., Fabian, A.~C., 
\& Blundell, K.~M.\ 2011, \mnras, 413, 1107 


\bibitem[Moss  \& Shukurov(1996)]{MS96} 	
 Moss, D., \& Shukurov, A. 1996, \mnras, 279, 229

\bibitem[Mulchaey \& Zabludoff(1998)]{MZ98} 
Mulchaey, J. S., \& Zabludoff, A. I. 1998, ApJ, 496, 73


\bibitem[Murgia(2003)]{M03} Murgia, M.\ 2003, PASA, 20, 19 

\bibitem[Murgia et 
al.(2011)]{MPM11} Murgia, M., Parma, P., Mack, K.-H., et al.\ 2011, \aap, 526, AA148

\bibitem[Nath(2010)]{N10} Nath, B.~B.\ 2010, \mnras, 407, 
1998 




\bibitem[O'Dea(1998)]{O98} O'Dea, C.~P.\ 1998, \pasp, 110, 
493 


\bibitem[Orienti 
\& Dallacasa(2010)]{OD10} Orienti, M., \& Dallacasa, D.\ 2010, 10th European VLBI Network Symposium and EVN Users Meeting: VLBI and the New Generation of Radio Arrays,  


\bibitem[Orienti et al.(2010)]{OMD10} Orienti, M., Murgia, 
M., \& Dallacasa, D.\ 2010, \mnras, 402, 1892 



\bibitem[Ostorero et al.(2010)]{OMS10} Ostorero, L., et al.\ 
2010, \apj, 715, 1071 

\bibitem[Ostriker \& McKee(1988)]{OM88} Ostriker, J.~P., \& McKee, C.~F.\ 1988, Reviews of Modern Physics, 60, 1 

\bibitem[Owsianik et al.(1999)]{OCP99} Owsianik, I., Conway, 
J.~E., \& Polatidis, A.~G.\ 1999, NewAR, 43, 669 






\bibitem[Perucho et al.(2011)]{PQM11} Perucho, M., Quilis, 
V., \& Mart{\'{\i}}, J.-M.\ 2011, \apj, 743, 42 



\bibitem[Polatidis 
\& Conway(2003)]{PC03} Polatidis, A.~G., \& Conway, J.~E.\ 2003, PASA, 20, 69 







\bibitem[Prandoni 
\& Seymour(2014)]{PS14} Prandoni, I., \& Seymour, N.\ 2014, arXiv:1412.6512 





\bibitem[Reynolds \& Begelman(1997)]{RB97} Reynolds, C.~S., \& Begelman, M.~C.\ 1997, \apjl, 487, L135 

\bibitem[Reynolds et al.(2001)]{RHB01} Reynolds, C.~S., 
Heinz, S., \& Begelman, M.~C.\ 2001, \apjl, 549, L179 


\bibitem[Reynolds et al.(2002)]{RHB02} Reynolds, C.~S., 
Heinz, S., \& Begelman, M.~C.\ 2002, \mnras, 332, 271 



\bibitem[de Ruiter et 
al.(2005)]{RPC05} de Ruiter, H.~R., Parma, P., Capetti, A., Fanti, R., Morganti, R., \& Santantonio, L.\ 2005, \aap, 439, 487 



\bibitem[Rybicki 
\& Lightman(1979)]{RL79} Rybicki, G.~B., \& Lightman, A.~P.\ 1979, New York, Wiley-Interscience, 1979.~393 p.,  



\bibitem[Schekochihin et al.(2005)]{SCK05} 	
 Schekochihin, A. A., Cowley, S. C., Kulsrud, R. M., Hammett, G. W.,
 \& Sharma, P. 2005, ApJ, 629, 139


\bibitem[Stage et al.(2006)]{SAH06} Stage, M.~D., Allen, 
G.~E., Houck, J.~C., \& Davis, J.~E.\ 2006, Nature Physics, 2, 614 


\bibitem[Stawarz et al.(2007)]{SCH07} Stawarz, {\L}., Cheung, 
C.~C., Harris, D.~E., \& Ostrowski, M.\ 2007, \apj, 662, 213 

\bibitem[Stawarz et al.(2008)]{SBM08} Stawarz, {\L}., 
Ostorero, L., Begelman, M.~C., Moderski, R., Kataoka, J., 
\& Wagner, S.\ 2008, \apj, 680, 911 

\bibitem[Tanaka et al.(2008)]{T08} Tanaka, T., et al.\ 
2008, \apj, 685, 988 



\bibitem[Taylor et al.(2000)]{TMP00} Taylor, G.~B., Marr, 
J.~M., Pearson, T.~J., \& Readhead, A.~C.~S.\ 2000, \apj, 541, 112 





\bibitem[Vikhlinin,  Markevitch,  \& Murray(2001)]{VMM01} 	
 Vikhlinin, A., Markevitch, M., \& Murray, S. S. 2001, ApJ, 549, L47


\bibitem[Wilson et al.(2000)]{WYS00} Wilson, A.~S., Young, 
A.~J., \& Shopbell, P.~L.\ 2000, \apjl, 544, L27 

\bibitem[Wilson et al.(2006)]{WSY06} Wilson, A.~S., Smith, 
D.~A., \& Young, A.~J.\ 2006, \apjl, 644, L9 


\bibitem[Yamazaki et 
al.(2004)]{YYT04} Yamazaki, R., Yoshida, T., Terasawa, T., Bamba, A., \& Koyama, K.\ 2004, \aap, 416, 595 




\bibitem[Yaji et al.(2010)]{YTI10} Yaji, Y., Tashiro, M.~S., 
Isobe, N., Kino, M., Asada, K., Nagai, H., Koyama, S., 
\& Kusunose, M.\ 2010, \apj, 714, 37 







\bibitem[Zanni et al.(2003)]{ZBR03} Zanni, C., Bodo, G., Rossi, P., Massaglia, S., Durbala, A., \& Ferrari, A.\ 2003, \aap, 402, 949 

 











































































%



































\end{thebibliography}
\end{document}